# Non-Thermal Resistive Switching in Mott Insulators

Yoav Kalcheim,[1] Alberto Camjayi,[2] Javier del Valle,[1] Pavel Salev,[1] Marcelo Rozenberg[3] and Ivan K. Schuller[1]

[1]Department of Physics and Center for Advanced Nanoscience, University of California-San Diego, La Jolla, California 92093, USA

[2]Departamento de Física, FCEyN, UBA and IFIBA, Conicet, Pabellón 1, Ciudad Universitaria, 1428 CABA, Argentina

[3]Laboratoire de Physique des Solides, CNRS, Univ. Paris-Sud, Université Paris-Saclay, 91405 Orsay Cedex, France

**Resistive switching can be achieved in a Mott insulator by applying current/voltage, which triggers an insulator-metal transition (IMT). This phenomenon is key for understanding IMT physics and developing novel memory elements and brain-inspired technology. Despite this, the roles of electric field and Joule heating in the switching process remain controversial. We resolve this issue by studying nanowires of two archetypical Mott insulators - $VO_2$ and $V_2O_3$. Our findings show a crossover between two qualitatively different regimes. In one, the IMT is driven by Joule heating to the transition temperature, while in the other, field-assisted carrier generation gives rise to a doping driven IMT which is purely non-thermal. By identifying the key material properties governing these phenomena, we propose a universal mechanism for resistive switching in Mott insulators. This understanding enabled us to control the switching mechanism using focused ion-beam irradiation, thereby facilitating an electrically driven non-thermal IMT. The energy consumption associated with the non-thermal IMT is extremely low, rivaling that of state of the art electronics and biological neurons. These findings pave the way towards highly energy-efficient applications of Mott insulators.**

The insulator-metal transition (IMT) in strongly correlated materials is a phase transition characterized by drastic changes in electrical properties, which may be accompanied by structural and magnetic transitions. The richness of this phenomenon has made it one of the most studied topics in condensed matter physics. The properties of IMT materials and their sensitivity to external stimuli make them promising for applications such as memory, selectors for ReRAM, optical switches and emulating brain functionalities.(*1*, *2*, *11–17*, *3–10*) Another advantage of these systems is that the IMT can be triggered by various perturbations, such as optical pumping or changes in temperature, pressure, strain and chemical doping. The most practical way of inducing the IMT in electronic devices is by applying electrical current or voltage.(*18*, *19*) Such devices can be very fast(*14*, *20*) and highly scalable.(*10*)

The mechanism behind the electrically-triggered IMT is currently under heavy debate in the scientific community. In materials where the IMT takes place as a function of temperature, such as $VO_2$ and $V_2O_3$, Joule heating due to the flow of electrical current is an obvious candidate for triggering the transition. It has been argued, however, that the electric field applied in this process may induce the transition without heating the material to its IMT temperature ($T_{IMT}$).(*21–26*) This has important consequences for the energy consumption, mechanical stability, and time scales involved in the switching process. Many studies have been carried out to address this issue, but contradictory results have been reported even for the same materials.(*20*, *21*, *34–36*, *25*, *27–33*) The greatest challenge in distinguishing between the effect of Joule heating and electric field is in determining the spatial distribution of current and temperature in a device. Because of the coexistence of metal/insulator domains close to the IMT, current flow can become highly non-uniform, resulting in inhomogeneous Joule heating and electric field profiles. For example, in quasi-2D $VO_x$ thin films, the current flow is filamentary which can dramatically alter the local temperature.(*19*, *37*) This hinders the determination of the mechanism behind the electrically-triggered IMT since it is unclear whether the transition occurs because the IMT temperature is locally attained under an applied field. In this work we study the electrically-triggered IMT in quasi-1D nanowires of two archetypical Mott insulators exhibiting an IMT: $VO_2$ and $V_2O_3$. Through accurate temperature calibration we find unambiguous evidence for both thermal and non-thermal resistive switching in both materials. Our work unveils the mechanism behind the electrically triggered IMT in Mott insulators and demonstrates how introducing defects can be used to control the switching mechanism.

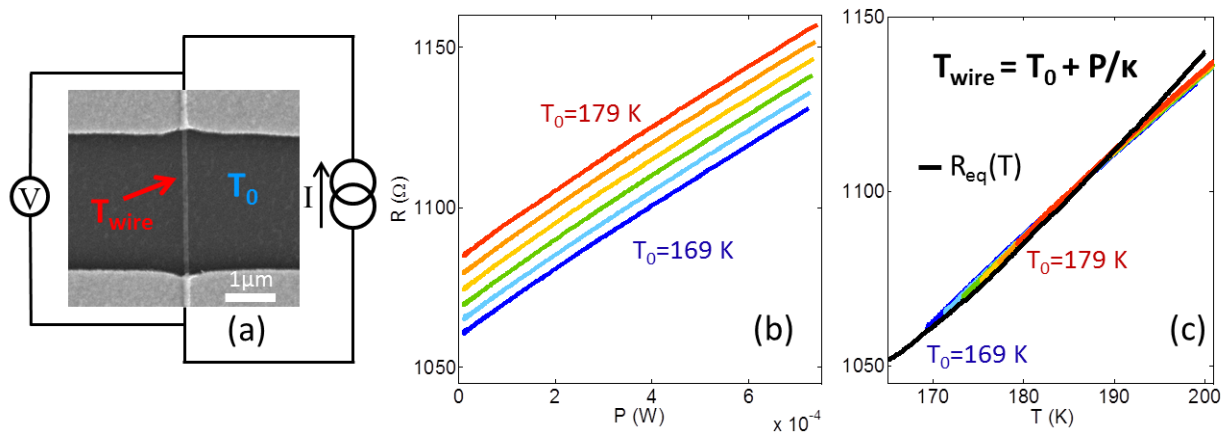

Fig. 1: **Determination of temperature change due to applied power in VOx nanowires. (a),** SEM image of a VO$_2$ nanowire contacted by two Ti/Au pads (top and bottom) and schematic of the measurement setup. $T_0$ and $T_{wire}$ are the temperatures of the substrate and the nanowire, respectively, which differ due to the applied power. **(b),** Resistance vs. power curves measured above the IMT temperature in the fully metallic state of a V$_2$O$_3$ nanowire. **(c),** Resistance vs. $T_{wire}$ (colored lines) derived from **(b)** using equation (2) with a single fitting parameter κ=24 μW/K for all curves (colored). The black curve shows the resistance vs. stage temperature at equilibrium, $R_{eq}(T)$. The horizontal T axis refers both to the temperature during the equilibrium (low power) measurement and $T_{wire}$.

## Power-to-temperature calibration

We employed advanced fabrication techniques to prepare high-quality nanowires of VO$_2$ and V$_2$O$_3$ (see scanning electron micrograph in Fig. 1(a)). (*38*) Our samples show extremely sharp 4-6 orders of magnitude resistance change across the IMT (supplementary Fig. S1(a) and (b)). The nanowire widths (~100 nm) are comparable to the size of insulator/metal domains(*39*) which is further supported by resistor network simulations (Supplemental Material section 1 – Fig. S1 (c-e)).(*38*) The advantage of using such quasi-1D structures is that spatial confinement suppresses filament formation during the electrically-triggered IMT. This greatly simplifies data interpretation and facilitates a clear distinction between thermal and electric field effects. Considering a homogeneous current distribution in the nanowire, the effect of Joule heating on the nanowire temperature ($T_{wire}$) can be quantitatively determined from a simple heat equation:

$$C \frac{dT_{wire}}{dt} = IV + \kappa(T_0 - T_{wire}) \qquad (1)$$

where I and V are the current and voltage applied to the nanowire, C is its heat capacitance, $T_0$ is the (measured) substrate temperature and $\kappa$ is the thermal coupling constant between the substrate and the nanowire. Equation (1) is strictly valid when power is evenly distributed and temperature variations along the nanowire can be neglected. This is the case in the single-phase homogeneous metallic and insulating states, above and below the IMT hysteresis. As will be shown subsequently, empirically, this equation still provides a very good approximation even when the two states coexist. In steady state $\frac{dT_{wire}}{dt} = 0$ and thus

$$T_{wire} = P/\kappa + T_0 \qquad (2)$$

where the power is P=IV. To measure $\kappa$ we performed a series of voltage vs. current measurements, V(I), at different $T_0$ corresponding to the metallic state. From this we derived resistance vs power curves - R(P), which show a trend of increasing R with P (see Fig. 1(b)). A similar trend is observed in resistance vs substrate temperature curves which were measured at thermal equilibrium [$R_{eq}(T)$] by applying minimal current/power to the sample and changing the stage temperature. Using equation (2) to relate P and $T_{wire}$, we successfully collapsed all the R(P) curves onto the $R_{eq}(T)$, as shown in Fig. 1(c). We note that all curves were collapsed using a single fitting parameter $\kappa$. For the nanowires discussed in this work, $\kappa$ ranged between 21-24 µW/K for the $V_2O_3$ devices and 45 µW/K for the $VO_2$ device (see Supplemental Material – section 2 for comparison of $\kappa$ values obtained in previous work).(38) Knowing these thermal coupling constants allows us to accurately quantify Joule heating and enables a reliable estimate of $T_{wire}$ from the measured parameters $T_0$ and P.

### Electro-thermal driven IMT in $VO_2$

To explore the thermal and electric field effects on the electrically induced IMT, current-controlled V(I) measurements were performed in the low temperature insulating state of the nanowires. For each V(I), the sample was initially cooled below the IMT hysteresis regime and then heated to $T_0$ to allow comparison with the heating branch of the $R_{eq}(T)$ measurement.

The R(P) curves for the $VO_2$ nanowire in the low temperature insulating regime are shown in Fig. 2(a). A series of abrupt resistance jumps is observed in the R(P) curves for $T_0>337$ K (light green curve). These are due to portions of the nanowire undergoing the IMT due to the applied current. These switching events occur with decreasing power as $T_0$ increases. Using only the κ value measured *above* the IMT, we used equation 2 to transform the R(P) curves measured at different $T_0$'s in the insulating state to $R(T_{wire})$. Fig 2(b) shows that the $R(T_{wire})$ curves collapse onto the $R(T_{eq})$, similarly to the collapse observed in the metallic state (see Fig. 1(a, c)). This collapse, extending well into the IMT, shows that electrical switching in this $VO_2$ nanowire can be accounted for solely by Joule heating. These findings are consistent with most previous reports which suggest that Joule heating plays a dominant role in the electrically induced IMT in $VO_2$.(*33*, *40*) The large jumps in the last section of some $R(T_{wire})$ curves are attributable to thermal runaway effects due to a current surge in the nanowire (see Supplemental Material – section 3).(*38*)

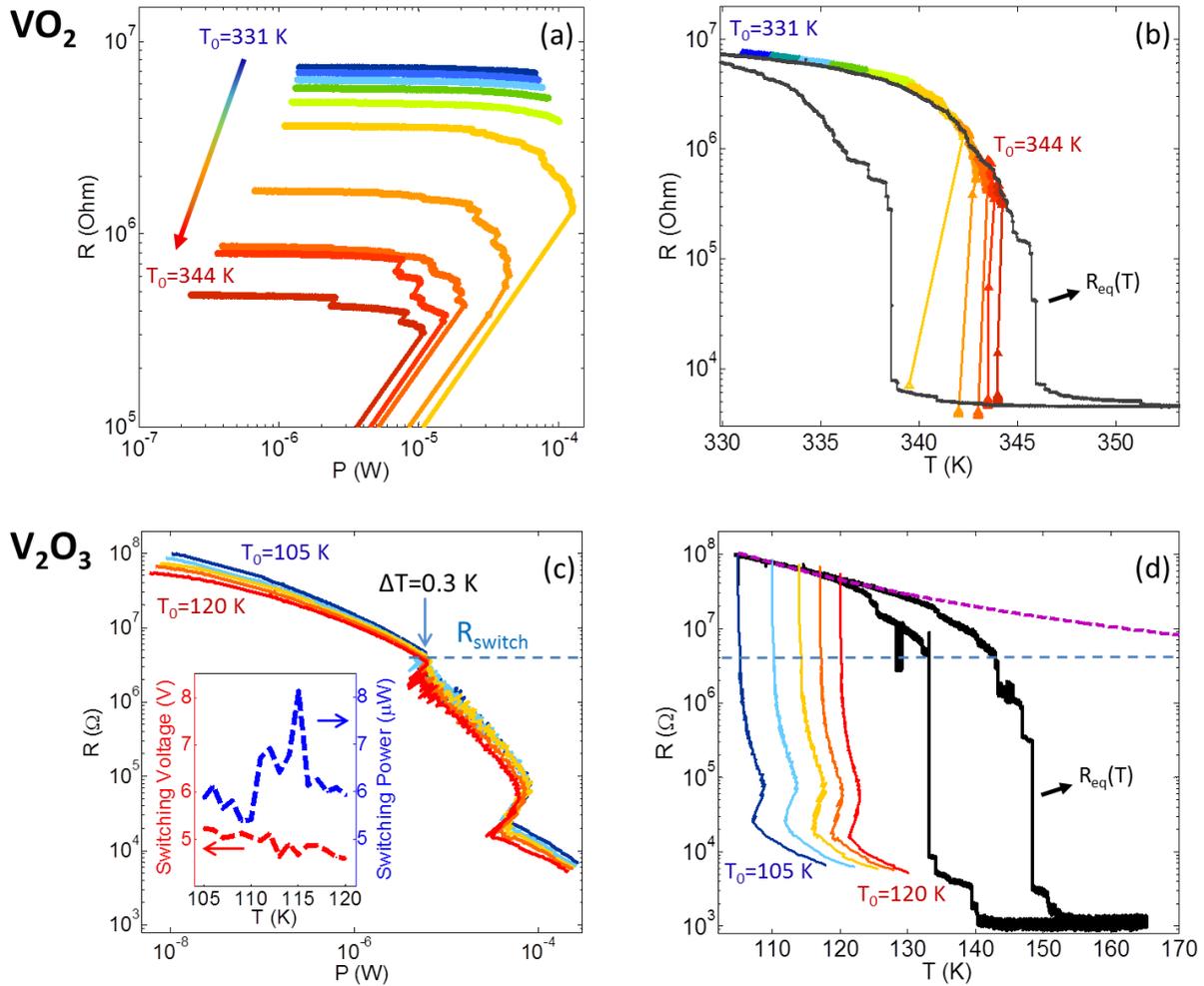

Fig. 2: **Thermal vs. non-thermal IMT switching. (a)** and **(c),** R(P) curves acquired at different substrate temperatures, $T_0$, corresponding to the insulating state of $VO_2$ and $V_2O_3$ nanowires, respectively. The IMT switching is manifested as abrupt resistance jumps. **(b)** and **(d),** $R(T_{wire})$ curves derived from the corresponding R(P) curves and superimposed on the $R_{eq}(T)$ (black). For the $VO_2$ nanowire in **(b)**, the excellent correspondence between $R(T_{wire})$ and $R_{eq}(T)$ signifies that the IMT is triggered thermally. For the $V_2O_3$ nanowire in **(d)**, the correspondence between $R(T_{wire})$ and $R_{eq}(T)$ is not observed. The IMT switching in **(c)** begins with only a 0.3 K increase in $T_{wire}$ regardless of $T_0$ (light blue arrow), indicating a negligible role of Joule heating. $R_{switch}$ (dashed blue line) is lower than the insulating state resistance of $R_{eq}(T)$ at any temperature. This is shown in **(d)** by the extrapolation of $R_{eq}(T)$ according to thermally activated resistance behavior with an activation energy of 60 meV (purple dashed line). The inset in **(c)** shows the switching voltage and power as functions of $T_0$.

## From thermal to non-thermal IMT

Considering the electrically induced IMT switching shown in Fig. 2(a,b) as a benchmark for a purely Joule heating driven transition, we show that $V_2O_3$ can exhibit qualitative differences. To assess the role of Joule heating, we follow the same procedure as before to estimate $T_{wire}$ from the measured R(P) in the metallic state and equation (2). The R(P) and corresponding $R(T_{wire})$ curves in the insulating state of a $V_2O_3$ nanowire are shown in Fig. 2(c,d), respectively. Initially, R(P) decreases *smoothly* by over an order of magnitude with increasing power up to ~6-8 µW. As can be deduced from comparison of the $R(T_{wire})$ to $R_{eq}(T)$, the decrease in resistance under applied voltage is not due to Joule heating since the dissipated power is too low. This smooth decrease is not attributed to the IMT either, as will be discussed subsequently. R decreases smoothly down to 3-5 MΩ, after which it drops abruptly by 20-30%. Interestingly, the switching resistance ($R_{switch}$) is nearly independent of $T_0$ (see Supplemental Material Fig. S6)(*38*) and is considerably lower than the equilibrium resistance in the fully insulating state at any temperature within the IMT (Fig. 2(d)). It is important to note that the resistance changes are reversible and the initial state is recovered once the applied current goes to back zero, thus ruling out irreversible resistive switching due to oxygen migration. These jumps therefore indicate that a significant portion of the nanowire has undergone switching into the metallic state. Surprisingly, the power just before switching (~7 µW) heats the nanowire by only ~0.3 K, despite a 15 K range in $T_0$. Moreover, the switching power is nearly independent of $T_0$, and varies in the range 5-8 µW (see inset in Fig. 2(c)). These findings are inconsistent with Joule heating-driven switching, for which the switching power would decrease with increasing $T_0$ by ~21 µW/K (as deduced from power-to-temperature calibration). On the other hand, the switching voltage ($V_{switch}$) decreases with increasing $T_0$, suggesting that the electric field is the primary driving force behind this electrically-triggered IMT (see Extended Data - section 3 for further discussion of Joule heating dynamics).

Interestingly, the resistive switching properties of different $V_2O_3$ nanowires showed some variability, most likely induced during the fabrication process. A systematic comparison of these properties allowed us to gain insight into the underlying mechanism of the electric-field-driven IMT. Fig. 3(a) shows $R_{eq}(T)$ curves of four $V_2O_3$ nanowires. The IMT in all four wires occurs over virtually the same temperature range. We find, however, that for any fixed temperature below the IMT hysteresis, the equilibrium insulating-state resistance varies between the nanowires. For instance, the resistance at 120 K ($R_{120K}$) in the heating branch of the $R_{eq}$(T) curves varies from 46 MΩ to 1100 MΩ. Since the nanowire geometries are nearly identical, these resistance variations are most likely due to small differences in defect concentrations. Defects may create in-gap trap states that release carriers by thermal activation, thereby decreasing the sample resistance in equilibrium(*41*) (see Supplemental Material - section 4 for discussion on activation energies).(*38*) Several recent studies have shown that deviations from perfect stoichiometry greatly affect the insulating state resistivity in $VO_2$ and $V_2O_3$.(*42, 43*) We thus use $R_{120K}$ as a proxy for the (inverse) defect density in the nanowires.

We show that defects play a crucial role in determining the non-equilibrium properties of $V_2O_3$ and, consequently, affect the mechanism by which the IMT takes place. To elucidate this, typical R(V) curves for the samples of Fig. 3(a) are shown in Fig. 3(b). For comparison, three R(V) curves were measured at different $T_0$ so that they have the same initial resistance (blue, green and red). Two important effects are observed as $R_{120K}$ decreases (defect concentration increases): (1) the initial slope of the smooth part of log[R(V)] curve becomes steeper and (2) the switching voltage ($V_{switch}$) decreases. In fact, the correlation between these non-equilibrium properties and $R_{120K}$ is virtually unaffected by varying $T_0$. Therefore, samples measured at lower $T_0$, unexpectedly, have *lower* $V_{switch}$ compared to samples measured closer to $T_{IMT}$, as shown in Fig. 3(b). For instance, the sample with $R_{120K}$=46 MΩ switches with only ~5.2 V (1.5 MV/m) at $T_0$=105 K, well below $T_{IMT}$. In contrast, the sample with $R_{120K}$=705 MΩ does not switch up to 12 V (3.4 MV/m) at $T_0$=138 K, despite being inside the phase coexistence (hysteresis) regime. The sample with the highest $R_{120K}$=1100 MΩ does not switch even with application of 20 V. The non-equilibrium properties are thus observed to be strongly dependent on the insulating state resistance, which is strongly affected by defects in the nanowire.

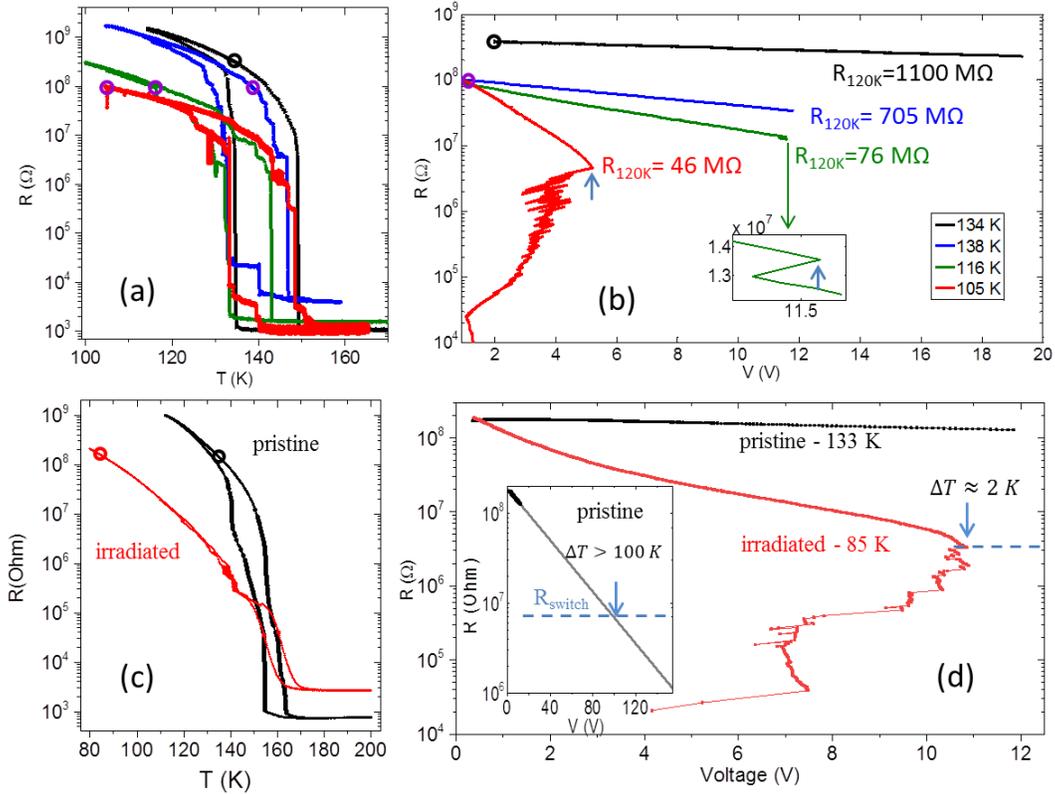

Fig. 3: **Influence of defects on the switching mechanism.** (a) Resistance vs. temperature curves of four $V_2O_3$ nanowires. Open circles denote $T_0$ for which the R(V) curves shown in (b) were acquired. (b), R(V) for the nanowires in (a). Samples with lower insulating state resistance, $R_{120K}$, exhibit steeper slopes of log(R) vs voltage, indicating enhanced carrier generation. For two samples, abrupt resistance jumps associated with the IMT are observed (light blue arrows). (c) $R_{eq}(T)$ before and after focused ion beam irradiation with a dose of $6.2 \cdot 10^{14}$ Ga ions/cm$^2$. Open circles denote $T_0$ for which the R(V) curves shown in (d) were acquired. (d) R(V) measured in the pristine and irradiated states. Similar to (b), the slope of log(R) vs voltage is considerably steeper in the irradiated state. The resistance reaches 2% of the initial value, $R_{switch}$=4 MΩ, followed by non-thermal switching with only ~2 K temperature increase. Assuming the same $R_{switch}$ for the pristine state, an extrapolation of R(V) predicts that switching occurs with over 110 V corresponding to a temperature increase of more than 100 K (inset in (d)).

To examine this further, we used focused Ga ion beam irradiation to systematically and controllably create defects in our samples. Fig. 3(c) shows the $R_{eq}(T)$ curves of a $V_2O_3$ nanowire before and after ion irradiation. Similar to the samples shown in Fig. 3(a), defects reduce the insulating state resistance but do not considerably affect the temperature range over which the IMT takes place. As in the non-irradiated samples, the R(V) characteristics show dramatic differences, due to the varying defect density. Fig. 3(d) shows two R(V) measurements acquired at $T_0$=133 K and 85 K in the pristine and irradiated states, respectively, so that the initial resistance is the same in both cases. For the pristine state, the resistance smoothly decreases to 73% of the initial value for the maximum voltage while the resistance of the irradiated sample smoothly decreases by almost two order of magnitude to ~2% of its initial value, followed by an abrupt resistance jump at $R_{switch}$=4 MΩ (light blue arrow). Just before switching, Joule heating in the nanowire corrresponds to a temperature increase of less than 2 K ($T_{wire}$<87 K). However, the first signatures of the IMT in $R_{eq}(T)$ occur at ~120 K, which would require an increase in $T_{wire}$ of ~35 K. Therefore, this switching cannot be explained by thermal effects. Qualitatively similar results were obtained also on irradiated $VO_2$ nanowires (see Supplemental Material – section 5),(*38*) thus corroborating the important role of defects in facilitating the electrically driven non-thermal IMT in Mott insulators.

**Mechanism for non-thermal switching and energy consumption**

We suggest a simple model to explain our experimental results based on destabilization of the Mott insulator phase by field-assisted excitation of carriers. First, we discuss the origin of the initial smooth decrease in resistance in response to the applied voltage as shown in Fig. 3(b,d). This decrease cannot be attributed to the IMT since switching even in a single domain would produce a resistance jump due to the low dimensionality of the nanowire, which is not observed. Furthermore, hysteretic behavior is not observed in the R(V) curves within the initial voltage-driven smooth decrease of R, even at temperatures *within the phase coexistence regime.* Similar voltage-driven decrease in resistance has been previously observed in Mott insulators(*20, 44*) and attributed to field-assisted carrier generation, known as the Poole-Frenkel effect.(*45*) The electric field reduces the energy barrier for excitation of trapped carriers from in-gap states into the conduction band (Fig. 4(b,d)). This increases the number of free carriers and decreases resistivity. In the weak field regime, the energy barrier decreases linearly with E, rather than the more commonly observed $E^{1/2}$.(*46*) This produces an exponential increase in carrier density with voltage, consistent with the observed exponential decrease of resistance with voltage (see Fig. 3(b)). The slope of these curves steepens with decreasing $R_{120K}$, indicating that higher defect densities increase the number of carriers which are excited for the same applied field.(*47*) As demonstrated in Fig. 2(d), the resistance just before switching ($R_{switch}$) is lower than the equilibrium resistance of the purely insulating phase, even if it is extrapolated to temperatures higher than that of the IMT. Therefore, we conclude that at $R_{switch}$ the carrier concentration is higher than that which can be attained in equilibrium conditions in the insulating state. This leads to destabilization of the insulating state through field-induced carrier doping, resulting in collapse of the Mott gap and emergence of the metallic state.

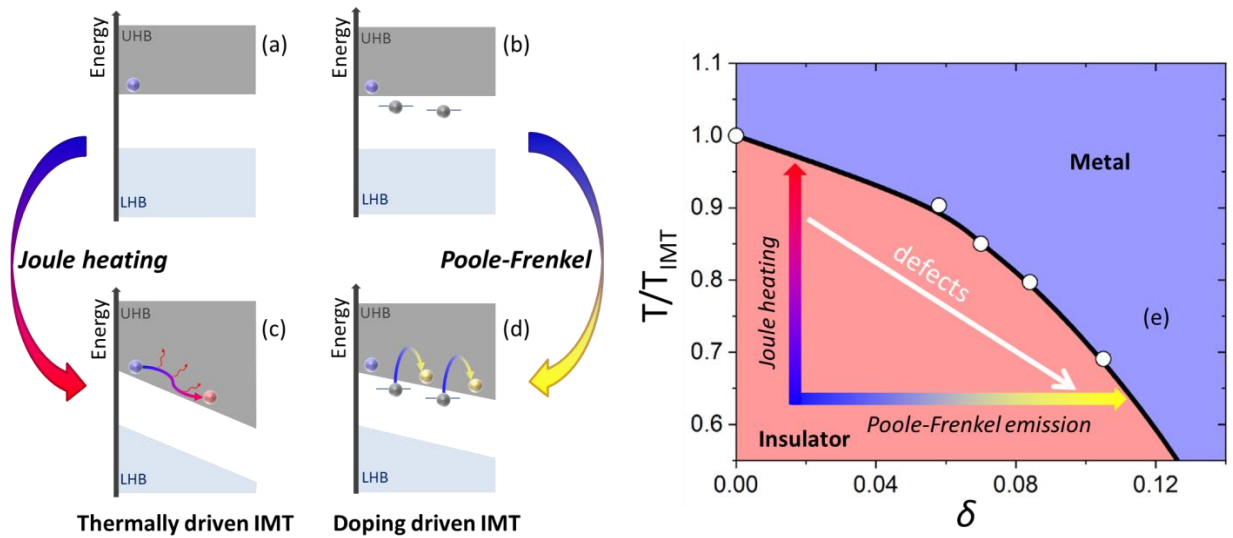

**Fig. 4: Two routes for resistive switching in a Mott insulator. (a-d)** Schematic representation of Joule heating and field-driven IMT switching in a Mott insulator. In a system with no defects **(a)**, only a small number of thermally activated carriers are present in the upper Hubbard band (UHB). Application of a strong electric field, depicted as strong band tilt **(c)**, accelerates these carriers causing Joule heating due to scattering. In a system with defects **(b)**, a moderate electric field promotes carriers from in-gap states into the UHB **(d)** causing the collapse of the Mott insulator state via a doping-driven IMT. **(e)**, Phase diagram of the single band Hubbard model based on DMFT calculations. The colored arrows schematically denote the two possible routes towards the metallic phase when applying electric field as discussed in **(a-d)**. Defects increase the effectiveness of the Poole-Frenkel emission and facilitate a doping driven non-thermal IMT.

In samples where the electric-field-assisted carrier generation is inefficient, the doping driven transition cannot be attained even for the highest applied voltages and the only possible route to induce the IMT switching is Joule heating past $T_{IMT}$ (Fig. 4(b,d)). This is most notable for the non-irradiated $VO_2$ nanowires (see Fig. 2(a) and supplementary Fig. S4)(*38*) and non-irradiated $V_2O_3$ nanowires with the highest insulating state resistances (black curves in Fig. 3(b) and 3(d)). For these samples, even up to the highest applied voltages, the resistance decreases very little compared to samples with high defect density. Therefore, the carrier density does not reach the critical value required to induce switching in the experimental voltage range. Since higher voltages may cause irreversible damage to the sample, we assess $V_{switch}$ by extrapolation. The inset to Fig. 3(d) shows a linear extrapolation of the log[R(V)] for the pristine nanowire to a resistance of 4 MΩ - the $R_{switch}$ value observed in the same sample after irradiation. Assuming that the linear dependence holds for higher voltages, switching would occur with over 110 V. This translates to Joule heating by over 100 K, which would bring the sample well above $T_{IMT}$. This indicates that thermal effects would play a major role in the electrical triggering of the IMT in this sample. We therefore conclude that there is a crossover in the switching mechanism from carrier generation to Joule heating. This crossover is controlled by the defect density, which determines the efficiency of field-assisted carrier generation, as depicted in Fig. 4. With higher defect densities, more carriers are generated with less applied field and power, thus enabling non-thermal switching. In contrast, low defect densities result in smaller carrier generation and Joule heating thermally triggers the IMT before the critical carrier concentration is attained.

We note that non-thermal switching in $VO_2$ requires higher irradiation doses compared to $V_2O_3$ (see Supplemental Material – section 5).(*38*) Even with these high doses, electric-field assisted carrier generation was not as effective as in $V_2O_3$, leading to higher $V_{switch}$ values in $VO_2$. This may be attributed to the significantly higher activation energy in $VO_2$ compared to $V_2O_3$, which could hinder field assisted carrier generation. Variation in activation energies may also contribute to the differences in switching characteristics as observed in the $V_2O_3$ nanowires (see Supplemental Material - section 4).(*38*) The strong dependence on specific sample properties can explain discrepancies regarding the nature of the IMT found in previous reports on the same materials.(*20, 21, 25, 48, 49*) These contradictory reports may be reconciled by considering differences in sample properties such as defect density and activation energy (see also Supplemental Material – section 2).(*38*)

To gain deeper insight into the physics of the electrically induced IMT, we performed dynamical mean field theory (DMFT) calculations for a single band Hubbard model. This model is considered the simplest framework for a qualitative description of the IMT in a strongly interacting electron system (Supplemental Material – section 6).(*38*) Motivated by our previous discussion, we assume that the main effect of the electric field is to promote charge carriers in the system, which is modeled by shifting the electric (chemical) potential to adjust the doping level. Starting from the insulator state, the resistivity $\rho(T)$ was calculated using the Kubo formula for various temperatures and doping levels to determine the regimes of insulating ($d\rho/dT<0$) and metallic ($d\rho/dT>0$) behavior. The goal is to qualitatively capture the observed initial smooth decrease of R down to $R_{switch}$. It is found that the IMT can be attained either by increasing temperature or by doping (see phase diagram in Fig. 4(e)). As detailed in the Supplemental Material, several important characteristics of the experimental switching are reproduced by the calculations.(*38*) We observe that small doping of just a few percent is sufficient to reach $R_{switch}$, where the system switches from insulating to metallic behavior. Moreover, we also observe that $R_{switch}$ is lower than the resistance in the insulating state in equilibrium, and has a weak dependence on temperature (see Supplemental Material - section 6). The ability of such a general and schematic model to capture important aspects of our experimental findings suggests that the proposed mechanism for the electric field driven IMT may apply to many Mott insulators beyond $VO_2$ and $V_2O_3$.

The ability to induce non-thermal switching without heating the device by more than a fraction of a Kelvin implies that this process can be highly energy efficient. Indeed, by applying nanosecond voltage pulses to the $V_2O_3$ nanowire discussed in Fig. 2(c,d), an upper bound of 5 fJ is found for the switching energy (Supplemental Material - Section 7). This presents an improvement in energy consumption of about 3 orders of magnitude over the Joule heating driven IMT and rivals resistive switching in state of the art memristors.(*50*) This has very important implications on the energetics of devices which utilize the IMT for memory or for emulating neural functionalities. Moreover, the time scales associated with the non-thermal

IMT may be much shorter than for the thermal IMT. For devices based on a thermal IMT, accumulation and dissipation of heat sets a lower bound on the operation time scales.(*51*) By largely avoiding heating during the switching process these time scales may be greatly reduced.

## **Conclusions and Outlook**

In this work we disentangle the effect of Joule heating and electric field on resistive switching in two archetypical Mott insulators - $VO_2$ and $V_2O_3$. We show clear evidence for thermal and non-thermal IMTs in both materials. Non-thermal switching occurs when field assisted carrier generation leads to critical doping levels where the Mott insulator is no longer stable. Several features of our experimental findings are reproduced by DMFT calculations, supporting the doping-driven IMT scenario. We find that defects are crucial for the non-thermal IMT, as they provide the charge reservoir for the doping process. To enhance their effect, large defect densities may be introduced into the sample during fabrication or by focused ion beam irradiation. This allows moderate fields to excite a large number of carriers with negligible heating. For low defect densities, carrier generation is unable to attain critical values, leaving Joule heating as the dominant switching mechanism. Our work demonstrates that, despite having two different ground states, $VO_2$ and $V_2O_3$ can be modified through defect engineering to exhibit a non-thermal IMT. This suggests that using controlled defects to induce non-thermal switching may be universally applicable to Mott insulators, thus paving the way towards a variety of highly energy efficient applications.

## **Acknowledgments:**

The synthesis and characterization aspects of this work were supported by the Vannevar Bush Faculty Fellowship program sponsored by the Basic Research Office of the Assistant Secretary of Defense for Research and Engineering and funded by the Office of Naval Research through Grant No. N00014-15-1-2848. The transport measurements and theoretical modelling were supported through an Energy Frontier Research Center funded by the U.S. Department of Energy, Office of Science, Basic Energy Sciences under Award No. DE-SC0019273. Part of the fabrication process was done at the San Diego Nanotechnology Infrastructure (SDNI) of UCSD, a member of the National Nanotechnology Coordinated Infrastructure (NNCI), which is supported by the US National Science Foundation under grant ECCS-1542148.

# Supplemental Material

## Methods:

### Sample preparation

**VO$_2$**: A 100 nm thick VO$_2$ film was grown on (012) oriented Al$_2$O$_3$ substrate by reactive rf magnetron sputtering. The growth was done in a 4 mtorr Ar/O$_2$ (92%-8%) atmosphere and at a substrate temperature of 470 °C. After the growth, the sample was cooled down at a rate of 12°C/min. X-ray diffraction showed textured growth along the (100) direction of VO$_2$. **V$_2$O$_3$**: 300 nm thick V$_2$O$_3$ films were grown on (012) oriented Al$_2$O$_3$ substrate by rf magnetron sputtering. The growth was done in a 8 mTorr Ar atmosphere and at a substrate temperature of 700 °C. After growth, the samples were thermally quenched at a rate of ~90 °C/min, which significantly improved the film's electrical properties.(*1*) X-ray diffraction showed textured along the (012) direction of V$_2$O$_3$.

For both materials, nanowires of 85-100 nm width were fabricated by e-beam lithography and reactive ion etching. Au (100 nm)/Ti (20 nm) electrodes were subsequently prepared by photolithography and e-beam evaporation to electrically contact 3-3.5 µm long nanowires.

### Transport measurements

Transport measurements were performed in a probe station using a current source pulse generator, nanovoltmeter and 20 GHz oscilloscope. To protect the sample against current surges during abrupt resistive switching, for some measurements a resistor was connected in series. Fast transport measurements to determine the energy consumption during non-thermal switching were performed using a bias tee to mix nanosecond pulses with a small DC current. The DC current was used to probe the resistance of the sample before and after the pulse. Due to the large impedance mismatch, a 50 Ω termination was added at the output of the pulse generator to avoid reflection.

Focused ion beam irradiation

Ion beam irradiation was performed in a FEI Scios DualBeam system using a focused ion beam (FIB) of 30 keV Ga ions. The etch rate was calibrated by measuring the dose required to etch the vanadium oxide film all the way to the substrate. For the dose used in this work ($6.2 \cdot 10^{14}$ ions/cm$^2$), the etching depth is estimated at ~1.4 Å, so that the film thickness is practically unaffected. A 5 μm by 5 μm region was irradiated by scanning the FIB at 1.5 pA with 50% overlap between successive locations to ensure a uniform irradiation dose across the entire nanowire.

## Section 1: R vs. T simulations

To determine whether switching is thermally driven or not one must take into account the current and temperature distribution in the device, which may be highly spatially inhomogeneous. To understand the reason for this, the differences between the electrically-driven and temperature-driven IMT should be considered. When the sample temperature is raised uniformly across the IMT, a VO$_x$ thin film will phase separate into metallic and insulating domains.(*2*) Metallic domains will nucleate and grow at the expense of insulating ones until the sample is fully metallic. If the typical size of such domains is small compared to the size of the device, a percolative transition will take place. In contrast, when driving the device with current, certain paths connecting the electrodes will have a lower resistance and they will attract more current and power than others. Eventually, a continuous path of metallic domains will form between the electrodes, manifesting as a sharp decrease in resistance and virtually the entire current will flow through a filament, while the rest of the sample will remain in the insulating state. With increasing power the filament will grow until the entire device is in the metallic state. During this process, the temperature in the device may be highly inhomogeneous and hard to measure or calculate. This makes it hard to determine if a transition can be induced without reaching the IMT temperature. Moreover, once a filament forms, the electric field on the device drops abruptly and only thermal effects from the heat injected through the filament will be observed subsequently. Therefore, when studying the importance of thermal effects on the IMT it is desirable to avoid filamentary conduction. This can be achieved by studying nanowires in which the domain size is comparable to the width of the nanowire.

For $V_2O_3$, thicknesses of 300 nm produce large domains, several hundreds of nanometers wide.(*3*) This large domain size is a result of high domain interface energy compared with thinner films.(*4*) The thickness of the VOx was selected so that the domain size is comparable or larger than the nanowire width. This manifests as large jumps in the $R(T)_{eq}$ curve, as shown in supplementary Fig. S1(a,b). When the size of the domains approaches the nanowire width the jumps become larger. From comparison with simulations of our $R(T)_{eq}$ curves we find that typically, 1-2 domains are present in a lateral cut of the nanowire (see (*5*) for simulation details). Thus, the inhomogeneous current and temperature distributions which may occur in 2D devices are avoided.

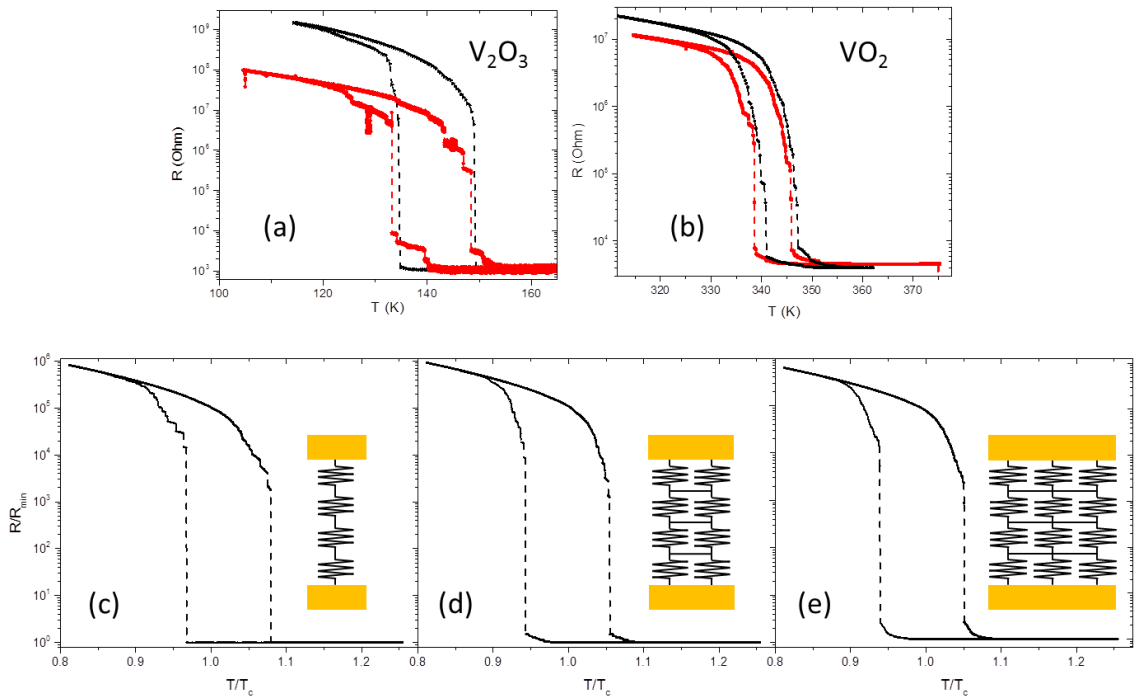

*Supplementary Fig. S1: (a) and (b) Measured resistance vs temperature curves for $V_2O_3$ and $VO_2$ nanowires. (c), (d) and (e) - Simulated R(T) curves for 1x20, 2x40, 3x60 resistor networks respectively. The biggest difference between the simulations is observed in the low resistance regime where the resistance jumps have the most impact on the shape of the curve. A one dimensional resistor chain has one abrupt jump to high resistances from the fully metallic state as observed experimentally in the black curve in (a) and in the simulation (c). The curves become smoother when the resistor network has more resistors per lateral cut.*

## Section 2 : Comparison to previous work

We have noticed that, in contrast to nanowires, continuous films such as the ones used in (6) show significantly less field assisted carrier generation, resulting in a dominantly Joule heating driven transition. This is probably due to the nanowire etching process which creates defects in the nanowire, similarly to the Ga irradiation process. Since most studies are conducted on continuous films, defect densities may be too low to observe non-thermal switching.

Another possible issue is that the thermal coupling constants found in the present study are almost an order of magnitude larger than those estimated in (*6*). There, $\kappa$ was deduced from the minimal power required to electrically maintain a highly conducting state below $T_C$ in a $V_2O_3$ junction. However, in this highly conducting state the current did not flow through the entire $V_2O_3$ sample. Therefore, this method produces a lower bound on $\kappa$ since maintaining the entire junction metallic requires significantly more power. This may explain why the conclusions of that study differ from the ones presented here, and highlights the importance of using 1D nanowires to accurately measure thermal coupling constants.

## Section 3: Simulation of Joule heating dynamics

We note that previous work on $V_2O_3$ devices suggests that a thermal runaway effect may inhibit the determination of $T_{wire}$ prior to a switching event in a DC measurement.(*6*) This effect is due to the decrease in resistance with temperature *in the insulating state*. For constant voltage, this may result in a positive feedback loop whereby increasing power decreases the resistance, which increases the power even further. To check if such an effect may apply to our case, we performed simulations using the experimentally measured parameters in the heat equation assuming that the voltage is constant even when the resistance goes down, due to parasitic capacitance. This yields the following equation:

$$C \frac{dT_{wire}}{dt} = \frac{V^2}{R(V, T_{wire})} + \kappa(T_0 - T_{wire}) \qquad (S1)$$

$R(V, T_{wire})$ is determined from R(V) measurements acquired at different temperatures with an activated behavior of the form (see section 3)

$$R(V, T_{wire}) = R_0(V) \exp(\Delta/k_B T_{wire}) \qquad (S2)$$

In our simulations we used the heat capacitance $C=1.46 \cdot 10^{-13}$ J/K derived from the literature value of (~35 J/mol·K).(7) However, since the heat capacitance can be absorbed into the time step, the stability analysis does not depend on its value, and all other parameters are measured. Our simulations show no thermal runaway effects when using the measured parameters, $V_0$=5.3 V, $\kappa$=21 µW/K and $\Delta$=60 meV. To reach instability by varying one of the parameters, the above values had to be changed to $V_0$=23 V (6.5 MV/m), $\kappa$=0.7 µW/K and $\Delta$=2000 meV respectively. Thus, thermal runaway effects in the insulating state are highly implausible in our samples and measurement conditions. This is further corroborated by the pulsed measurements discussed in the last section of the extended data. We note that in (6) a large contribution from Joule heating is observed most easily at low T where the switching fields are significantly larger than 6.5 MV/m so that thermal runaway effects may play an important role.

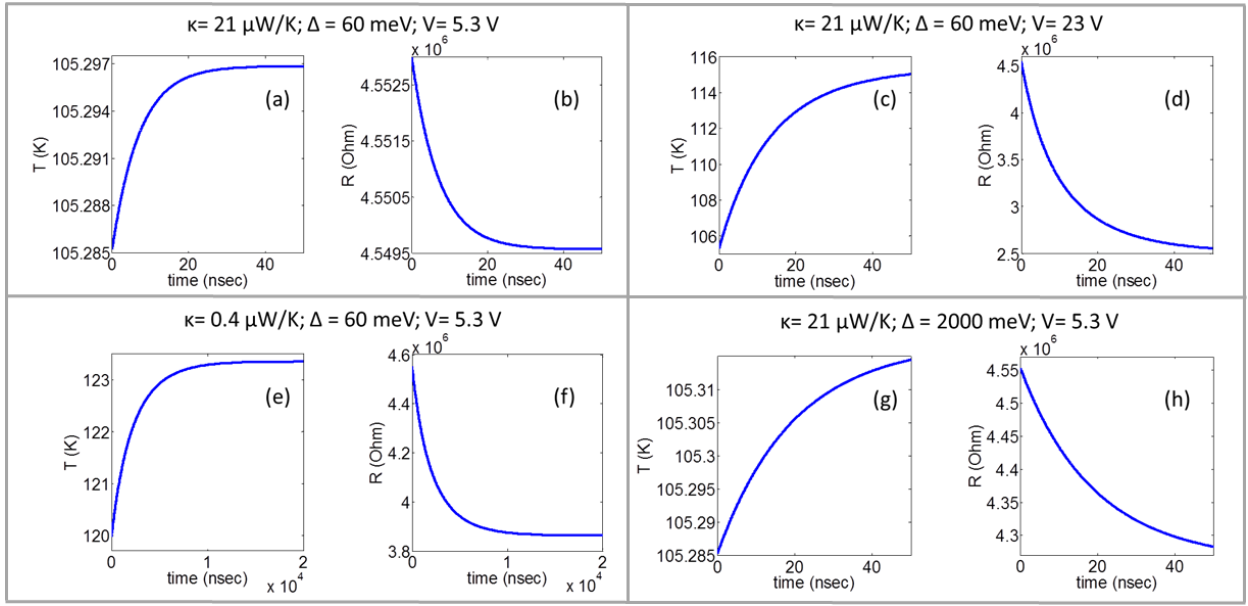

*Supplementary Fig. S2: Simulated temporal evolution of the temperature and resistance following a voltage increase from 5.2 V to the specified value. This analysis shows the range of parameters for which no thermal runaway effects are attained for the heat equation S1. The results for the measured parameters are shown in (a, b). These parameters are well within the stability range, indicating that thermal runaway effects within the insulating state are highly implausible in our samples.*

During thermal runaway, a large portion of the sample becomes metallic and, once the sample has cooled down, the resistance drops abruptly to that of the cooling curve at the measurement temperature. We note that even the formation of metallic domains does not necessarily result in thermal runaway effects. When the sample is in the fully insulating state, small jumps are often observed which only change the resistance by several percent, indicating that the system is robust against thermal runaway effects even when metallic domains form. Runaway effects are observed only when resistance jumps represent a large fraction of the total resistance. Then, the current surge is significantly larger and may lead to positive feedback. For the nanowire geometry, this occurs when few insulating domains are left in the sample, which can explain the jumps in $R(T_{wire})$ down to the cooling curve values, as shown in Fig. 2 of the main text for the $VO_2$ nanowire.

## Section 4: Activation energies

Due to the strong dependence of the resistance on applied field for $V_2O_3$ the activation energies for the various samples were derived from resistance vs temperature curves acquired at a fixed voltage of 4 V (see Fig. S2). The fit was performed to the low temperature part of the heating curve where no resistance jumps are observed so that the sample was completely in the insulating state at all fitted temperatures. As long as the sample is in the fully insulating state (no jumps observed in the V(I)), the activation energies are not sensitive to the choice of voltage. For $VO_2$ the fitting was performed to the R(T) curve acquired with constant current. For the $10^{-7}$ A applied during this R(T), the deviation with respect to a constant voltage R(T) curve are negligible, as can be seen from the small change in resistance with voltages up to 14 V for the pristine sample shown in supplementary Fig. S4.

Interestingly, we find that the activation energies vary considerably more for $V_2O_3$ nanowires than for $VO_2$, as do the switching characteristics. Lower activation energies correlate with enhanced Poole-Frenkel emission and lower switching voltage and power. Both defect density and activation energy seem to play a role in determining switching characteristics. See further discussion in section 5.

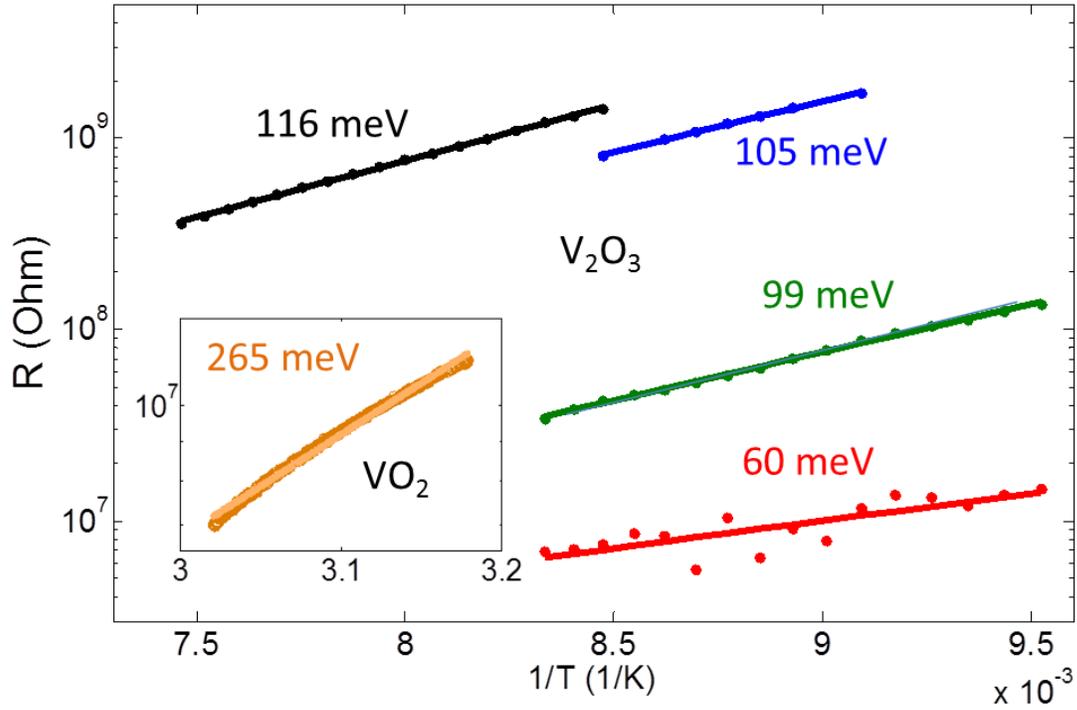

*Supplementary Fig. S3: Activation energies for the $V_2O_3$ and $VO_2$ samples discussed in the main text according to $R(T) = R_0 \exp(\Delta/k_B T)$. The color coding of the $V_2O_3$ samples corresponds to Fig. 3(a) of the main text.*

## Section 5: Irradiation effect in $VO_2$

A similar irradiation procedure to the one described in the main text was performed for $VO_2$ nanowires (see methods). Two R(V) curves are shown in supplementary Fig. S4 for a nanowire irradiated with $6.2 \cdot 10^{15}$ ions/cm² and a pristine nanowire. The R(V) for the irradiated sample is measured at 310 K, well below the IMT, whereas the R(V) for the pristine sample is measured at 335 K, within the phase coexistence regime (see arrows in supplementary Fig. S4). The R(V) curve of the irradiated $VO_2$ nanowire shows a steeper slope compared to that of a pristine sample. As in $V_2O_3$, this shows that defects increase field assisted carrier generation. A subsequent abrupt resistance drop occurs at ~13 V in the irradiated sample, signifying an electrically induced IMT. The power right before the switching is 23 μW which is equivalent to heating by ~0.5 K (see details of power calibration procedure in Fig. 1 of main text). However, the onset of the IMT is 15 K higher than $T_0$ ($T_{onset}$=325 K), thereby ruling out a thermally induced IMT. Despite being in the phase coexistence regime and applying similar voltage and power, the pristine sample does not show any switching.

Interestingly, when compared to $V_2O_3$, field assisted carrier generation is considerably less efficient in the case of $VO_2$ as can be deduced from the smaller smooth change in resistance for the same applied fields. This, in turn, results in higher electric fields required for switching. For instance, $V_{switch}$ values of a few Volts were observed in $V_2O_3$ even in pristine samples, whereas all measured $VO_2$ samples showed $V_{switch}$ values of over 12 V below the coexistence regime. This may be related to the differences in the magnitude of the activation energies observed in both cases. For $V_2O_3$ activation energies were between 60 meV and 116 meV whereas in $VO_2$ the activation energy is considerably larger (~265 meV). In $V_2O_3$ we observe a trend of increasing effectiveness of carrier generation and correspondingly lower $V_{switch}$ for lower activation energies (see discussion in section 2 of extended data). This trend is consistent with the case of $VO_2$, indicating that the activation energy also plays an important role in determining the effectiveness of field assisted carrier generation and, consequently the IMT switching characteristics. An open question remains, whether defect engineering may facilitate a higher degree of control over these important materials properties.

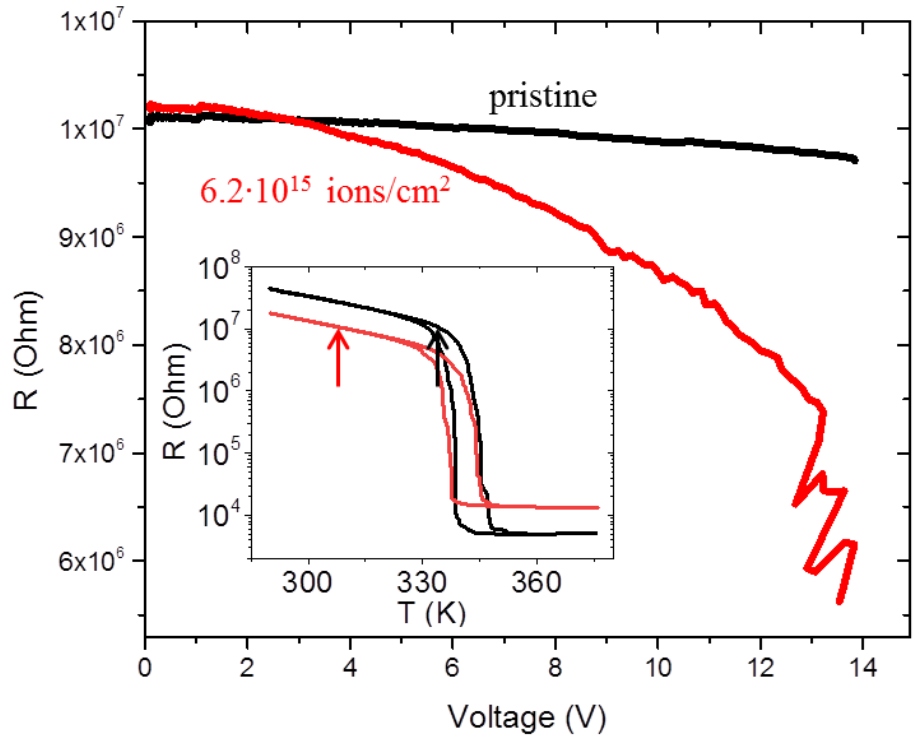

*Supplementary Fig. S4: Resistance vs voltage measurement of a pristine $VO_2$ nanowire (black) and an ion irradiated $VO_2$ nanowire (red - $6.2 \cdot 10^{15}$ Ga ions/cm$^2$). The non-irradiated sample shows a small ~3% decrease in resistance at the highest V. The irradiated sample shows a considerably larger decrease in R with V and subsequent switching (abrupt resistance drops). The estimated increase in temperature at the first resistance drop is 0.5 K, whereas the IMT onset in equilibrium occurs more than 20K above the measurement temperature. This shows the non-thermal nature of the resistive switching, as is the case in $V_2O_3$. The inset shows the R(T) for both nanowires along with the corresponding measurement temperatures.*

## Section 6: Theoretical Modelling

Dynamical Mean Field Theory (DMFT)(*8*) provides a non-perturbative method to study the interplay between correlation effects and electron banding and has in particular produced insights into the Mott metal-insulator transition. DMFT theory maps a lattice problem onto a quantum impurity model, a finite-size system coupled to a non-interacting bath of electrons, plus a self-consistency condition. The simulations were performed using the continuous-time quantum Monte Carlo (CTQMC)(*9*), which samples a diagrammatic expansion of the partition function in powers of the impurity-bath hybridization. In this method the on-site Hamiltonian is solved exactly, and the coupling to the bath is treated by a perturbation expansion. We solve for the single band Hubbard model with an antiferromagnetic self-consistency condition,(*8*) which captures both the antiferromagnetic and paramagnetic solutions. For the non-interacting bath electrons we adopt a semi-circular density of states of half-bandwidth W = 1. Similar models have been employed for exploring the electrically driven IMT in previous theoretical studies.(*10*, *11*)

The calculations require solving the system across the thermal and carrier doping transition, in the very low doping regime. Furthermore, high precision is required for the calculation of the conductivity from Kubo formula. This is enabled by using CTQMC to obtain solutions of the Hubbard model at finite temperatures. We consider the model at intermediate correlation strength (U/W=1.7, where U is the Coulomb repulsion and W is the half-bandwidth), and examine the system at half-filling (i.e. one carrier per site) where it is, like $V_2O_3$, an antiferromagnetic insulator. Supplementary Fig. S5(a) displays the resistivity as a function of T where a second order IMT is observed, marked by a change in the sign of the slope of the $\rho$(T) curve. In many Mott insulators, once the metallic state sets in, a structural transition occurs simultaneously,(*12*) creating a large discontinuous decrease in resistance. For simplicity, the structural transition, along with the discontinuity in resistivity, is not included in this calculation since we are only interested in studying the destabilization of the insulating state with temperature/doping. We assume that the main effect of the electric field is to promote charge carriers in the system, which are described by doping the Mott insulator via an electric (chemical) potential.

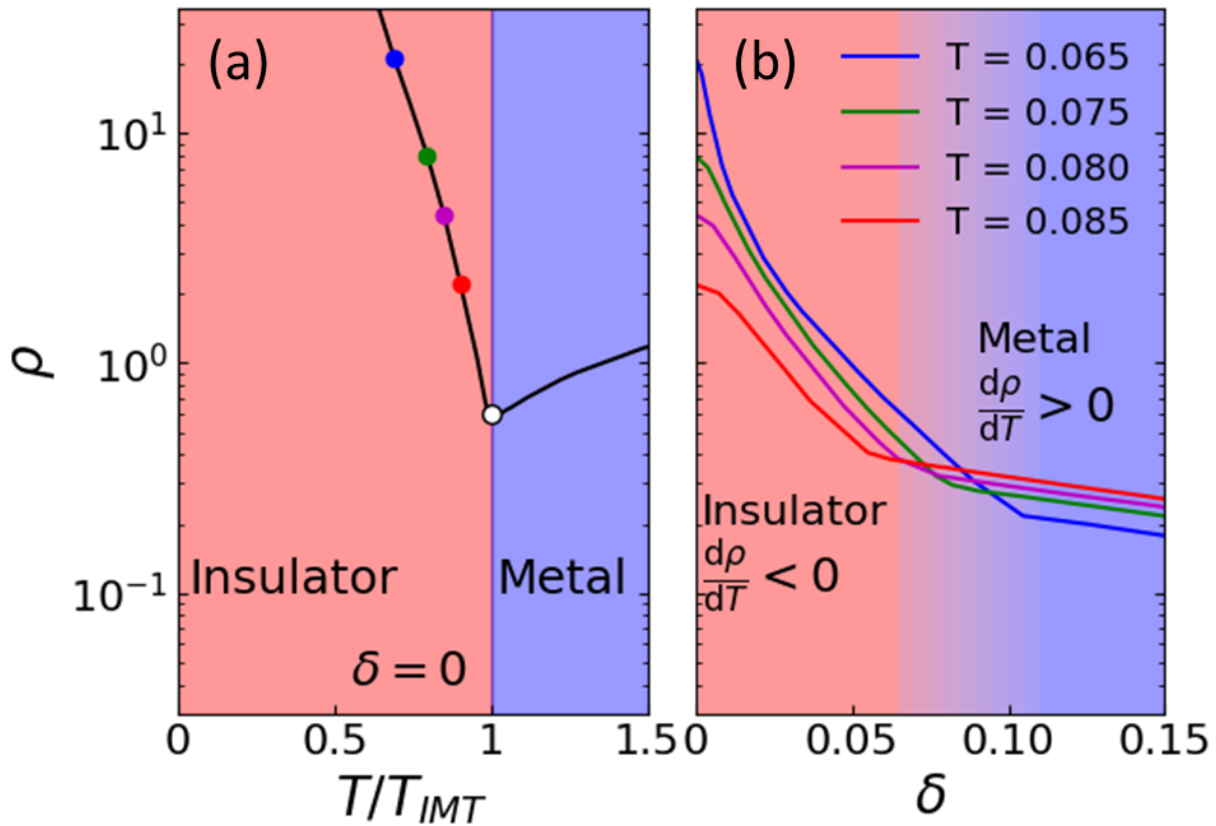

*Supplementary Fig. S5: (a) resistivity as a function of temperature in the temperature driven insulator to metal transition at zero doping for a single band Hubbard model. (b) Doping driven transition for different starting temperatures showing a change from insulating (dρ/dT<0) to metallic behavior (dρ/dT>0). While ρ has a strong dependence on temperature at zero doping, the dependence of ρ at the doping driven IMT is only weakly T dependent, as observed experimentally.*

Despite the simplicity of the model, several important experimental features are qualitatively reproduced by the calculation. Supplementary Fig. S5(b) shows the calculated resistivity as a function of doping for different T across the IMT. For low doping $d\rho/dT<0$ as expected in the insulating phase and $\rho$ is strongly temperature dependent. As $\delta$ increases the temperature dependence weakens until the resistivity curves cross each other and at high doping $d\rho/dT>0$ (metallic state) for the entire modelled temperature range. Despite the strong dependence of $\rho(\delta=0)$ on T, the resistivity at the critical doping ($\rho^*$) is only weakly temperature dependent. This is consistent with the experimental observation that $R_{switch}(T_0)$ is nearly independent of $T_0$, while larger temperature dependence is observed for $R(T)_{eq}$ (see supplementary Fig. S6).

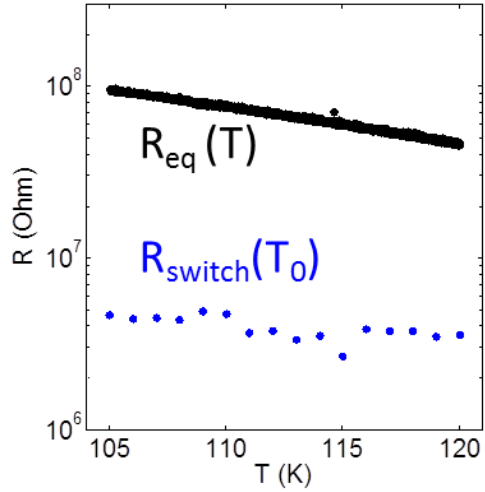

*Supplementary Fig. S6: $R(T)_{eq}$ and switching resistance $R_{switch}(T_0)$ for the nanowire with $R_{120K}=46$ M$\Omega$ discussed in the main text. The horizontal axis refers both to the temperature during the equilibrium measurement and the different stage temperatures $T_0$ during V(I) sweeps.*

The calculations also show that for all T, $\rho^*(\delta>0) < \rho^*(\delta=0)$. This is also observed experimentally as shown by the comparison between $R_{switch}$ and the extrapolation of $R(T)_{eq}$ in the insulating state to temperatures above $T_{IMT}$ (see Fig. 2(d) in main text). This can be explained by considering that an increase in δ from zero at $T_{IMT}$ drives the system into the metallic state and to a lower ρ. Now, to reach the new transition temperature for the non-zero doping state, the temperature has to be lowered below $T_{IMT}$. Since dρ/dT is positive in the metallic state, the resistance at the transition point for δ>0 has to be lower than $\rho(T_{IMT},\delta=0)$.

## Section 7: Switching energy measurements

In this section we discuss the switching energy of the nanowire shown in Fig. 2(c,d) in the main text. Pulses were applied in the fully insulating state so that power would be uniformly distributed along the nanowire. To observe a change in resistance after switching, the sample had to be in the coexistence regime where some metallic domains would remain stable after applying the pulse. Before applying pulses, R(V) measurements were performed at $T_0=112$ K from which $V_{switch}=5.5$ V was determined (see supplementary Fig. S4). Due to the high sample impedance, time resolved measurements could not be performed. Instead, the resistance was measured in the DC limit by applying a continuous current of 10 nA while 6 ns voltage pulses of varying amplitudes were applied, using a bias tee (see also methods). Twenty pulses were applied with total amplitudes of 4.5 V and 4.9 V with no observable change in sample resistance. Next, a single 5.3 V pulse was applied and switching was observed, consistent with the $V_{switch}=5.5$ V found in DC measurements. The switching resulted in a decrease of ~15% of the sample resistance. Using the known V(t) during the pulse and steady state R(V) curve we derived the power input to the nanowire as a function of time. By integrating the power over the duration of the pulse, the total energy delivered to the nanowire was found to be 5 fJ. Using the literature value for the heat capacitance of $V_2O_3$ (~35 J/mol·K)(*7*), this corresponds to a ~120 mK increase in the nanowire temperature. Since the resistance change for such a small temperature increase is extremely small, no thermal instabilities are expected to occur in this case. This supports the thermal stability simulations and rules out thermal runaway effects in the insulating state, as suggested in a previous study.(*6*) We note that the shortest pulse duration we could apply was 6 ns, so that 5 fJ is an upper bound for the energy required to induce the transition. In fact, IMT transition times as fast as several picoseconds have been observed in pump-probe experiments for $V_2O_3$.(*13, 14*) If this transition could be electrically triggered with a similar voltage over picoseconds instead of nanoseconds, the switching energy may be reduced by at least two orders of magnitude.

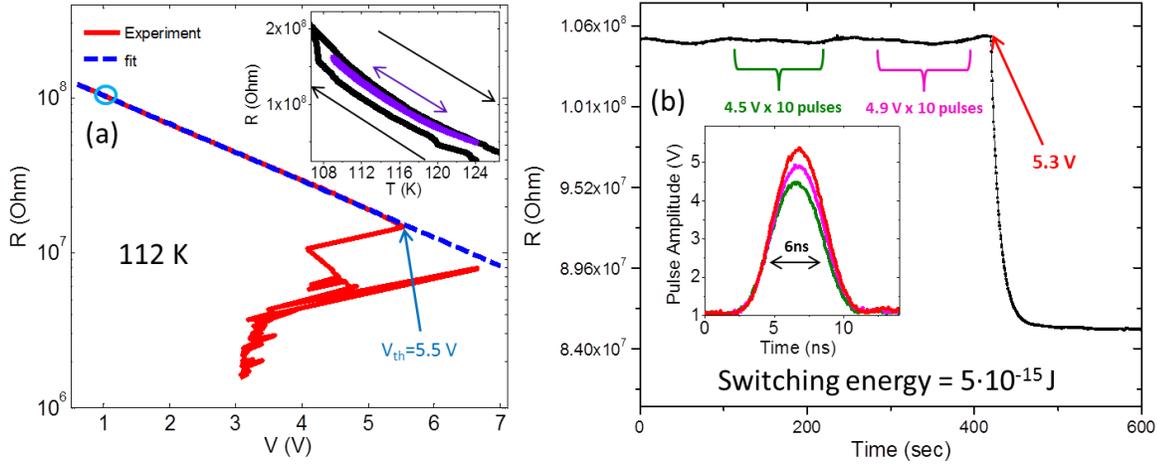

*Supplementary Fig. S7: (a) R(V) curve acquired at 112 K showing switching at 5.5 V. This curve is used to determine the resistance during the voltage pulse shown in (b). The inset to (a) shows $R(T)_{eq}$ curves acquired along the full IMT loop (black) and along a minor loop (purple curve - 109 K → 124 K → 109 K). The minor loop shows no hysteresis. (b) R(time) acquired with I=10 nA (equivalent to light blue circle in (a) at ~1 V) while 6 ns voltage pulses with varying amplitudes are applied to the sample. The various pulses are shown in the inset. 20 pulses of 4.5 V and 4.9 V did not show any resistance change while the first 5.3 V pulse triggered a large change in resistance, in accordance with the threshold voltage measured in (b). The slow relaxation of resistance observed after switching is due to RC discharge.*

To compare the switching energy to what is expected from a Joule-heating driven transition we measured the lowest temperature required to observe an IMT in part of the sample in equilibrium. We performed an R(T) minor loop following the sequence 109 K → 124 K → 109 K (see purple curve in inset to Fig. S4). No jumps were observed during this measurement and the sweeps up and down in temperature were repeatable. We thus conclude that for a Joule heating scenario the sample must be heated from $T_0$=112 K to at least 124 K to induce the IMT in a portion of the nanowire. Based on the heat capacitance we find that ~1.5 pJ must be imparted to the nanowire to reach this temperature. This value would be even larger if heat dissipation during the pulse were also considered. Moreover, the latent heat associated with the transition for the entire nanowire is ~7 pJ. These energies are nearly three orders of magnitude larger than the energy imparted to the nanowire during the pulse, indicating a negligible role played by Joule heating.

These results are in stark contrast to the case of the $VO_2$ nanowire discussed in Fig. 2(a,b) of the main text. For example, at 334 K, almost 5 times higher voltage (25 V) and 75 times more power (~150 μW equivalent to $\Delta T$~3.5 K) can be continuously applied without switching the $VO_2$ nanowire, despite being very close to the IMT.